# Reinforced magnetic properties of Ni-doped BiFeO$_3$ ceramic

**J. S. Hwang, Y. J. Yoo and Y. P. Lee***

*Department of Physics, Hanyang University, Seoul 04763, Korea*

**J.-H. Kang**

*Dept. of Nano and Electronic Physics, Kookmin University, Seoul 136-702, Korea*

**K. H. Lee and B. W. Lee**

*Hankuk University of Foreign Studies, Yongin 449-791, Korea*

**S. Y. Park**

*Advanced Institutes of Convergence Technology, Seoul National University, Suwon 443-270*

Multiferroic materials attract considerable interest because of the wide range of potential applications such as spintronic devices, data storage and sensors. As a strong candidate for the applications among the limited list of single-phase multiferroic materials, BiFeO$_3$ (BFO) is a quite attractive material due to its multiferroic properties at room temperature (RT). However, BFO is widely known to have large leakage current and small spontaneous polarization due to the existence of crystalline defects such as oxygen vacancies. Furthermore, the magnetic moment of pure BFO is very weak owing to its antiferromagnetic nature. In this paper, the effects of Ni$^{2+}$ substitution on the magnetic properties of bulk BFO have been investigated. BFO, and BiFe$_{0.99}$Ni$_{0.01}$O$_3$, BiFe$_{0.98}$Ni$_{0.02}$O$_3$ and BiFe$_{0.97}$Ni$_{0.03}$O$_3$ (BFNO: Ni-doped BFO) ceramics were prepared by solid-state reaction and rapid sintering, and analyzed by structural and magnetic-property measurements. The leakage current density was measured at RT by using a standard ferroelectric tester. All the Ni-doped BFO exhibit the similar rhombohedral perovskite structure (*R3c*) to that of BFO. The magnetic properties of Ni-doped BFO are much enhanced with respect to BFO prepared at the same conditions, since the enhanced ferromagnetic interaction is caused by the Fe/Ni coupling.

Email: yplee@hanyang.ac.kr
Fax: +82-2-2281-5572

# I. INTRODUCTION

Multiferroics are known as materials that have coupled electric and magnetic order, which result from the coexistence of ferromagnetism and ferroelectricity [1, 2]. The control of magneto-electric properties is in significant interest for applications in advanced multifunctional devices such as magnetic storage media and spintronics [3, 4]. During past decade, some oxide materials have been mentioned as multiferroics follow as $BaTi_2O_4$, $BiFeO_3$, $BiMnO_3$, $CuO$, $LuFe_2O_4$, $YMnO_3$ and so on [5-10].

Among the multiferroics, $BiFeO_3$ (BFO) is good candidate material for multiferroics due to the fact that ferroelectricity and antiferromagnetism coexist at room temperature (RT). The structure of BFO has rhombohedrally-distorted perovskite structure (with the space group *R3c*), and its ferroelectric (FE) and antiferromagnetic (AFM) (G-type) behavior occur below the Curie temperature ($T_C$) ∼ 1103 K and Neel temperature ($T_N$) ∼ 643 K, respectively [11]. The magnetic structure of BFO has G-type AFM ordering with an incommensurate spiral spin structure having a long periodicity of 62 nm [12]. The antisymmetric Dzyaloshinskii-Moriya (D-M) exchange interaction produces a weak ferromagnetic (FM) moment (∼ 0.02 $\mu_B$/Fe) due to spin canting of the AFM sublattices [13-15]. Moreover, BFO has a high leakage current attributed to the highly-volatile Bi ions [16]. It is difficult to improve the FE property, which constricts its application fields. The early studies have reported that, by replacing atoms at the *A* or the *B* sites in the perovskite structure, the multiferroic properties could be improved [17, 18]. For example, substitution with diamagnetic elements such as Ca and Pb ions at the *A* sites makes the enhancement of magnetization [19, 20]. In these compounds, the value of the spontaneous magnetization depends on the kind of substituting elements. Similarly, the replacement with Ti ions at the *B* sites exhibits a decrease in leakage current due to filling of the oxygen vacancies in BFO [21]. Recently, the enhancement of the FM property in 5% Ni-doped BFO ceramic has been realized [22].

In this paper, we report variation of the FM and the FE properties by changing Ni doping concentration. Our studies find that the FM properties of BFNO are greatly improved, which provides the possibility of multiferroic device for the practical applications.

## II. FABRICATION AND MEASUREMENT METHOD

Polycrystalline ceramics of BFO and BFNO were prepared by using the standard solid-state reaction method and rapid sintering. It has been reported as difficult to obtain a single-phase BFO [23]. Rapid sintering has been considered to be helpful in suppressing the formation of undesirable phase [24]. The stoichiometric amounts of high-purity (99.99% or better) $Bi_2O_3$, $Fe_2O_3$, and NiO powders were thoroughly mixed and subsequently calcined in air at 450 °C for 24h, with intermediate grinding for homogenization. Then, the samples were directly put into an oven, which was heated up to 800 °C and sintered in air for 20 min. The sintered disks were taken out from the oven and cooled to RT within several min. The crystallinity of samples was analyzed by using powder x-ray diffraction (XRD) (Rigaku Miniflex) using CuK$_a$ radiation at RT. The magnetic hysteresis loop at RT was performed with vibrating-sample magnetometer [VSM: VSM-7404, Lakeshore]. The FE properties and leakage current were obtained using the standard FE tester (Radiant-RT66B).

## III. RESULTS AND DISCUSSION

The crystallinity of BFO and BFNO are investigated by XRD as in Fig. 1. The small impurity peaks around $2\theta = 30°$ (marked by stars), corresponding to $Bi_{12}(Bi_{0.5}Fe_{0.5})O_{19.5}$ phase, are detected for BFO [25]. The observed diffraction peaks of BFNO exhibit the rhombohedral perovskite structure (*R3c*) similar to BFO. No other impurities are observed in Figs. 1(b), 1(c) and 1(d), reflecting that the $Ni^{2+}$ substitution for the Fe site does not affect the crystal structure of BFO. In addition, we notice few change in the lattice parameters between BFNO and BFO, as shown in

Fig. 1. Lattice constants *a* and *c* of BFNO are slightly increased according to Ni doping concentration due to the fact that the ionic radius of $Ni^{2+}$ ion (radius = 0.069 nm) is larger than that of $Fe^{3+}$ ion (radius = 0.0645 nm).

Figure 2 presents the magnetization vs. magnetic-field (*M-H*) hysteresis loops at RT of BFO and BFNO, which was performed at the maximum applied magnetic field of $H_{appl}$ = 10 kOe. The *M-H* curves of BFO exhibit linear ones with small remnant magnetization and coercive field, suggesting the AFM behavior with weak ferromagnetism [26]. On the other hand, the *M-H* loops of all BFNO ceramics show nearly-saturated FM behavior. When the Ni concentration in BFO becomes to be higher, the value of remnant magnetization ($M_r$) is enhanced. $M_r$ is 0.05, 0.05 and 0.07 emu/g for the Ni concentration of 0.01, 0.02 and 0.03, respectively, which is much larger than that of BFO (0.007 emu/g). This is because of the oxygen vacancies by $Ni^{2+}$ substitution, which lead to suppression of the spiral spin structure, thus providing the enhanced magnetization [27]. However, the oxygen vacancies are known well to alter slightly the magnetization of BFO [28, 29]. Therefore, the enhanced magnetization of the Ni-doped ceramics in perovskite BFO is induced by the D-M exchange interaction [13-15]. Taking into account this fact, one can suggest that Fe/Ni coupling leads to the FM interaction which is responsible for the weak magnetism. Additionally, the magnetization can be enhanced owing to the breakdown of the balance between antiparallel sublattices [30]. The local lattice distortion, which is caused by the $Ni^{2+}$ substitution, gives rise to the FM super-exchange interaction with disturbance of the antiparallel magnetic ordering, and therefore causes the enhancement of the magnetization of BFNO ceramics [31]. At the same time, it is clearly seen that the BFO sample exhibits unsaturated magnetization. The BFNO ceramics also present the weak ferromagnetism and the lack of saturation. This can be understood by competing contributions of the AFM (linear term) and the FM (hysteresis term) behaviors of BFO.

To investigate the electric properties of BFO and BFNO ceramics, we measured the FE hysteresis (*P-E* curve) at RT using a FE tester with electric-field sweep from -6 to 6 kV/cm [Fig. 3]. The BFO ceramics show nearly-saturated polarization at 5.2 kV/cm with small double remnant polarization ($P_r$) of 0.005 μC/cm$^2$. Even though a BFNO ceramic exhibits large $P_r$ of 0.663 μC/cm$^2$ (Ni doping of 3%) compared to BFO, the *P-E* loop has the rounded feature of lossy one originating from the existence of space-charge carriers related to the increased oxygen-vacancy concentration. The FE off-centering distortion arises because the Ni-doped BFO possesses a higher degree of noncentrosymmetry structure as compared with the pure BFO [32]. For the Ni-doped ceramics, unclosed hysteresis loops are observed, which indicate that the breakdown electric field is very low [33].

The lossy-loop polarization can be relevant to high leakage current due to the oxygen vacancy by the $Ni^{2+}$ substitution. As shown in Fig. 4, the leakage current of Ni-doped ceramics increases with respect to the pure BFO. The typical values of the current densities at 2 kV turn out to be $8.0\times10^{-9}$, $1.5\times10^{-4}$, $5.5\times10^{-4}$ and $4.1\times10^{-6}$ A/cm$^2$, respectively, for 0%, 1%, 2% and 3% Ni concentration. In the pure BFO, the oxygen stoichiometry creates the valence fluctuation of Fe ions (+3 to +2 state) and in turn it enhances the electric conductivity more [34]. In BFNO ceramics ($Ni^{2+}$ < 3%), the doping of $Ni^{2+}$ in place of $Fe^{3+}$ ions is compensated by the substituted lower-valance $Ni^{2+}$ and thus the oxygen vacancies are created. The charge compensation mechanism in BFNO is the source of oxygen vacancies, which lead to higher density of free carriers and increase the conductivity [35].

## III. CONCLUSIONS

The BFO and the Ni-doped BFO ceramics were synthesized using the standard solid-state reaction and rapid sintering technique. The XRD patterns show formation of the rhombohedrally-distorted perovskite structure for all samples. In addition, the lattice constant of BFNO is found to be slightly larger than that of BFO. Enhanced magnetic properties are observed in the Ni-doped

BFO ceramics. It is evident that the local lattice distortion caused by the $Ni^{2+}$ substitution is effective in improving the magnetic properties, by suppressing the spiral spin structure in BFO. The magnetization of Ni-doped BFO turned out to be greatly improved when the Ni doping concentration is 3%.

## ACKNOWLEDGEMENT

This work was supported by the ICT R&D program of MSIP/IITP, Korea (13-911-01-101).

Figure Captions.

Fig. 1. XRD patterns of BFO and Ni-doped BFO ceramics.

Fig. 2. Magnetization versus magnetic field for BFO and Ni-doped BFO ceramics at RT.

Fig. 3. *P-E* curves of BFO and Ni-doped BFO ceramics, (a) BFO, (b) BiFe$_{0.99}$Ni$_{0.01}$O$_3$, (c) BiFe$_{0.98}$Ni$_{0.02}$O$_3$, (d) BiFe$_{0.97}$Ni$_{0.03}$O$_3$.

Fig. 4. Leakage current for BFO and Ni-doped BFO ceramics.

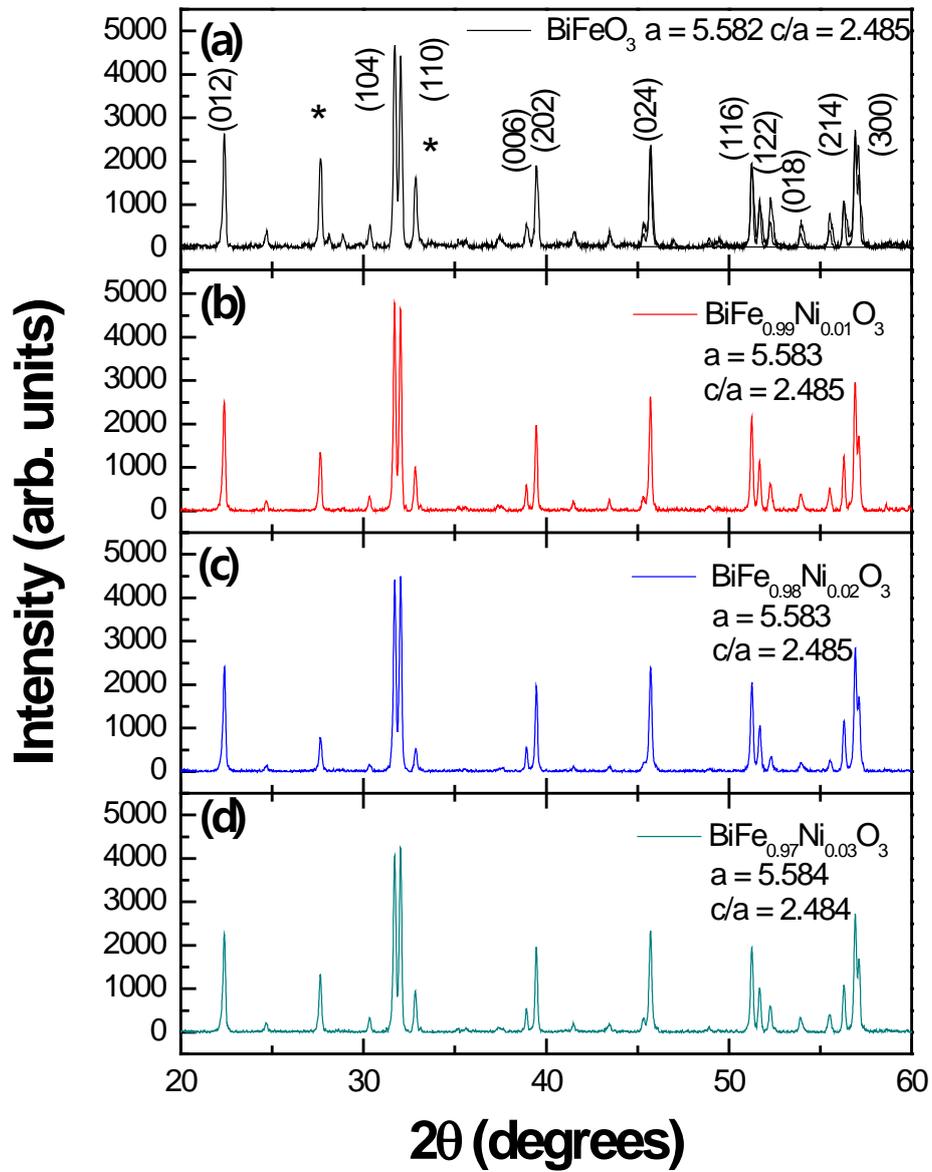

Fig. 1. XRD patterns of BFO and Ni-doped BFO ceramics.

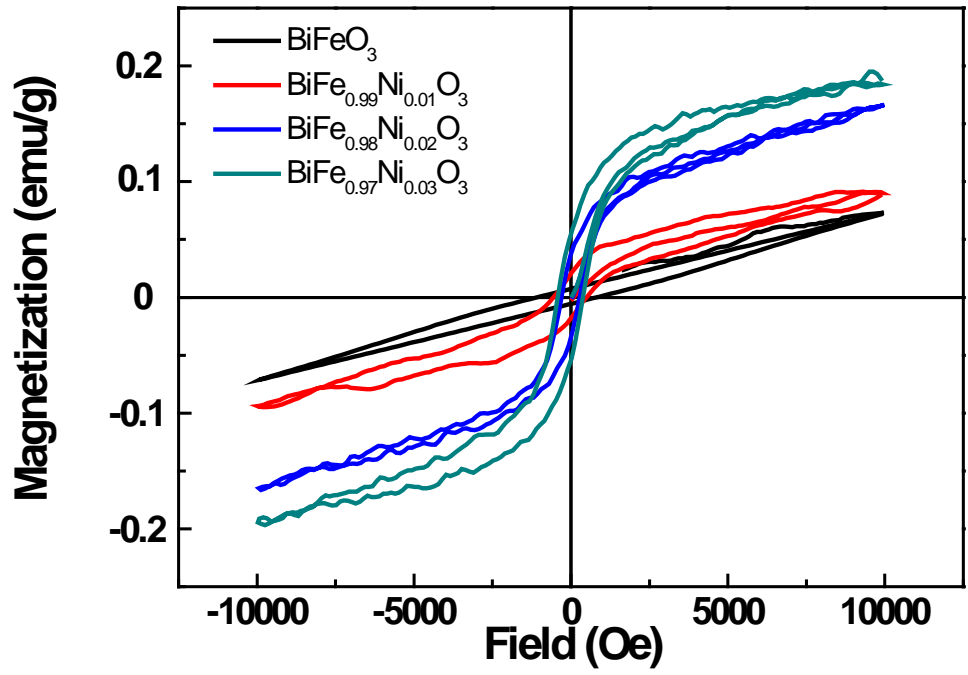

Fig. 2. Magnetization versus magnetic field for BFO and Ni-doped BFO ceramics at RT.

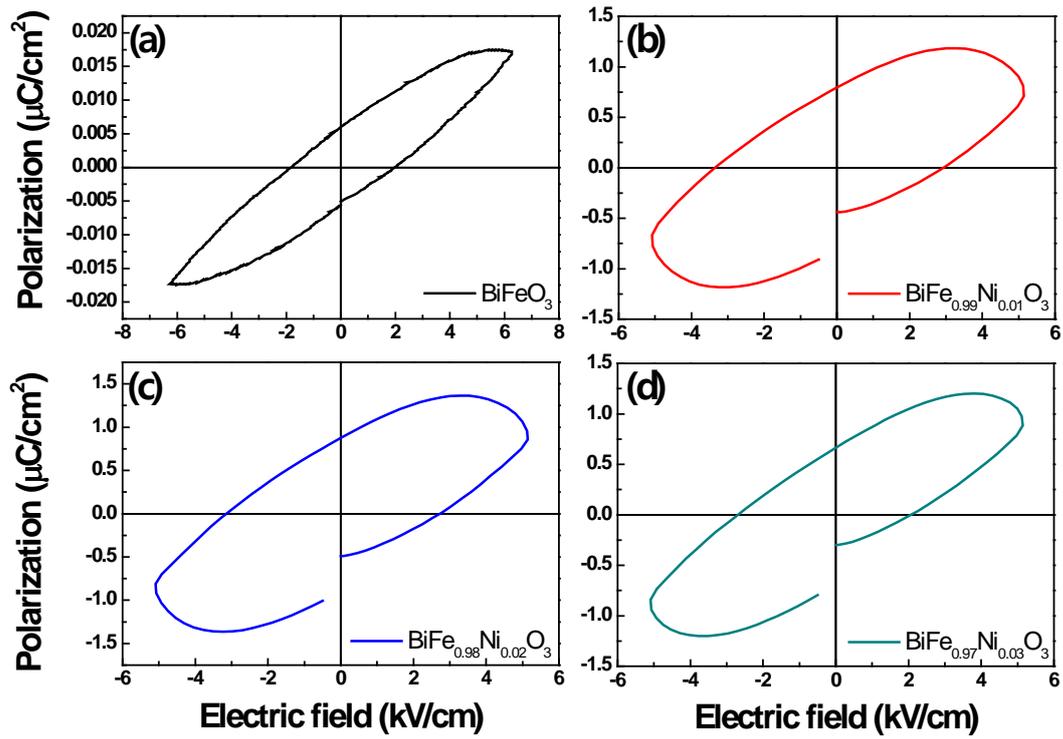

Fig. 3. *P-E* curves of BFO and Ni-doped BFO ceramics, (a) BFO, (b) BiFe$_{0.99}$Ni$_{0.01}$O$_3$, (c) BiFe$_{0.98}$Ni$_{0.02}$O$_3$, (d) BiFe$_{0.97}$Ni$_{0.03}$O$_3$.

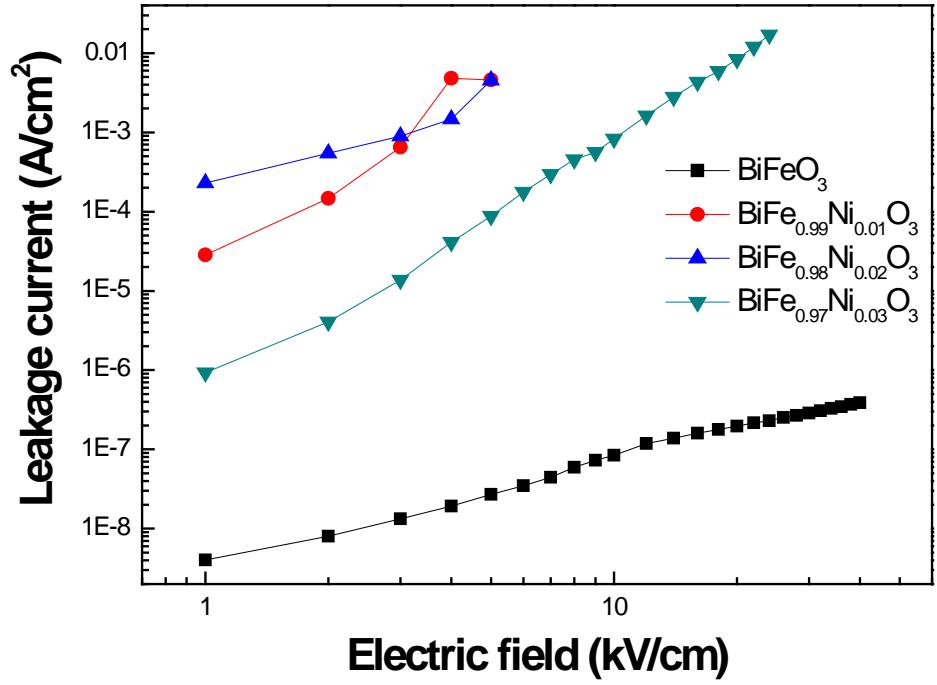

Fig. 4. Leakage current for BFO and Ni-doped BFO ceramics.